\documentclass{aa}  
\usepackage{graphicx}
\usepackage{txfonts}

\usepackage[hidelinks]{hyperref}
\usepackage{amsmath}
\usepackage{amssymb}	
\usepackage[T1]{fontenc}
\usepackage{ae,aecompl}
\usepackage{xcolor}
\usepackage{multirow}
\usepackage{tabularx}
\usepackage{threeparttable}
\usepackage{natbib}
\usepackage{adjustbox}
\usepackage{makecell}
\bibpunct{(}{)}{;}{a}{}{,} 

\usepackage{siunitx}

\begin{document}

   \title{Fe K$\alpha$ and Fe K$\beta$ line detection in the NuSTAR spectrum of the ultra-bright Z-source Scorpius X-1}

   \author{S. M. Mazzola\inst{1,2}, R. Iaria\inst{2}, T. Di Salvo\inst{2}, A. Sanna\inst{1}, A. F. Gambino\inst{2}, A. Marino\inst{2,4}, E. Bozzo\inst{3}, C. Ferrigno\inst{3}, A. Riggio\inst{1}, A. Anitra\inst{2} \and L. Burderi\inst{1}
    }

   \institute{Dipartimento di Fisica, Universit\`a degli Studi di Cagliari, SP Monserrato-Sestu, KM 0.7, Monserrato, 09042 Italy
         \and
   Dipartimento di Fisica e Chimica - Emilio Segrè, Universit\`a di Palermo, via Archirafi 36 - 90123 Palermo, Italy
    \email{simonamichela.mazzola@unipa.it}
         \and
        Department of Astronomy, University of Geneva, Ch. d'Ecogia 16, 1290, Versoix (Geneva), Switzerland
        \and 
        INAF/IASF Palermo, via Ugo La Malfa 153, I-90146 - Palermo, Italy
             }

\date{ }

 
  \abstract
  {Low-mass X-ray binaries hosting a low-magnetised neutron star, which accretes matter via Roche-lobe overflow, are generally grouped in two classes, named Atoll and Z sources after the path described in their X-ray colour-colour diagrams. 
   Scorpius X-1 is the brightest persistent low-mass X-ray binary known so far, and it is the prototype of the Z sources.
   }
   {We analysed the first \textit{NuSTAR} observation of this source to study its spectral emission exploiting the high statistics data collected by this satellite. 
   Examining the colour-colour diagram, the source was probably observed during the lower normal and flaring branches of its Z-track. We separated the data from the two branches in order to investigate the evolution of the source along the track. 
   }
   {We fitted the 3-60 keV \textit{NuSTAR} spectra using the same models for both the branches. We adopted two description for the continuum: in the first case we used a blackbody and a thermal Comptonisation with seed photons originating in the accretion disc; in the second one, we adopted a disc-blackbody and a Comptonisation with a blackbody-shaped spectrum of the incoming seed photons. A power-law fitting the high energy emission above 20 keV was also required in both cases.
   }
   {The two models provide the same physical scenario for the source in both the branches: a blackbody temperature between 0.8 and 1.5 keV, a disc-blackbody with temperature between 0.4 and 0.6 keV, and an optically thick Comptonising corona with optical depth between 6 and 10 and temperature about 3 keV. Furthermore, two lines related to the K$\alpha$ and K$\beta$ transitions of the He-like \ion{Fe}{xxv} ions were detected at 6.6 keV and 7.8 keV, respectively. A hard tail modelled by a power law with a photon index between 2 and 3 was also required for both the models.
 }
{}

   \authorrunning{S. M. Mazzola et al.}

  \titlerunning{Fe K$\alpha$ and Fe K$\beta$ line detection in Sco X-1 \textit{NuSTAR} spectrum}
  

   \maketitle
%

\section{Introduction}
In the Low-Mass X-ray Binaries harbouring neutron stars (hereafter
NS-LMXBs) a weakly magnetised neutron star (NS) accretes matter from a low-mass ($< 1$ M$_\odot$) companion star via Roche-lobe overflow.
A sub-classification of NS-LMXBs is based on the spectral and timing variability of the sources \citep{Hasinger_89}. The pattern traced by a single source in its X-ray colour-colour diagram (CD) or hardness-intensity diagram (HID). Thus we distinguish the so-called Atoll-class (with luminosity $\sim$ 0.01-0.1 of the Eddington luminosity $L_{\rm Edd}$) and Z-class (luminosity close to $L_{\rm Edd}$) systems.
The CD of the Z-sources shows the typical three branches pattern, in which we identify the horizontal branch (HB) at the top of the track, the normal branch (NB) in the middle and flaring branch (FB) at the bottom \citep[see][]{Hasinger_89}. The evolution of an individual source along the Z-track occurs in a timescale of few days and it is (probably) driven by the variability in the mass accretion rate $\dot{M}$ \citep{Hasinger_90}.

The power spectrum of a Z-source shows quasi periodic oscillations (QPOs), i.e. low-amplitude X-ray modulation with frequencies between 5-1250 Hz. According to the standard scenario, QPOs are due to the interaction between the weak ($<10^{10}$ G) magnetic field of the NS and transient blobs of accreting matter in the innermost region of the accretion disc \citep[see][ for a review]{VanderKlis_89,vanderklis_06a}. The QPOs frequencies could be compatible to the beat between the NS spin frequency and the Keplerian frequency of the blobs \citep{Alpar_85} and result in a modulation in the mass accretion rate $\dot{M}$ chargeable to the X-ray intensity variation \citep{Lamb_85}.
Since the highest frequency oscillations, the so-called kHz QPOs \citep[see e.g.][]{Strohmayer_96,vanderKlis_00,Jonker_00,VanderKlis_06b}, occur near the orbital frequencies of matter in the inner accretion disc, a model dependent constraint on the mass and the radius of the NS can be inferred assuming a stable orbital motion around the NS with radius between the NS surface and the innermost stable circular orbit (ISCO$ = 6 GM/c^2$). Under these hypotheses, it is possible to find an upper limit on the observable frequency at the ISCO in the range 1000-1250 Hz, assuming a NS with mass $M_{\rm NS}=2 M_{\odot}$ \citep{zhang_97,Miller_98,Miller_16}.

In the continuum of the X-ray spectra of NS-LMXBs we identify, in general, a soft thermal component due to the blackbody emission from the NS and/or the accretion disc, and a hard component due to the Comptonisation of soft photons from a hot electron corona located (probably) in the inner region of the system, around the NS or above the inner disc \citep[see e.g.][]{Dai_10,pintore_15}. Furthermore, the spectra of these sources show often a reflection component, originated from direct Compton scattering of the Comptonised photons outgoing the hot corona with the cold electrons in the top layers of the inner accretion disc; in most of the cases it can show the so-called Compton hump above 10 keV \citep[see e.g.][]{Egron_13,Miller_13, Ludlam_17,Coughenour_18,Ludlam_20,Ludlam_21}. The reflection component shows also some discrete features due to the fluorescence emission and photoelectric absorption by heavy ions in the accretion disc. The strength of the reflection is mainly indicated by the presence of a strong broad (FWHM up to 1 keV) emission line from Fe atoms between 6.4 and 6.97 keV (Fe-K region), identified with the K$\alpha$ radiative transition of iron at different ionisation states \citep[e.g.][]{iaria_19,Iaria_16,Papitto_13,Sanna_13,Miller_13,disalvo_09,Iaria_09,Iaria_07}.
These features most likely originate in the region of the accretion disc closer to the compact object, where matter is rapidly rotating and reaches velocities up to a few tenths of the speed of light \citep[see e.g.][and references therein]{Mazzola_19,DiSalvo_15}. Hence, the whole reflection spectrum is believed to be modified by transverse Doppler shifts, Doppler broadening, relativistic Doppler boosting and gravitational redshift, which produce the characteristic broad and skewed line profile \citep{fabian_89}. On the other hand, there are some Z-sources where reflection spectral components are absent \citep[as for example the source GX 5-1, see][]{bhulla_19,homan_18,jackson_09}, suggesting different geometries or accretion flow properties or different metallicities.

The Z-sources present also a hard power-law component predominant above 20 keV 
\citep[e.g.][]{Reig_16,paizis_06,iaria_01a,disalvo_01a} which strength is usually related with the position of the source in the CD, being highly significant in the HB up to disappear in the FB \citep[see e.g.][]{disalvo_00,dai_07}. The origin of the hard tail is still matter of debate; it may originate by the Comptonisation in a hybrid thermal/non-thermal electron corona \citep[see e.g.][]{farinelli_05,poutanen_98}, or in a mildly relativistic bulk motion of matter close to the compact object \citep[e.g.][]{farinelli_08,psaltis_01}.

In this work, we show the study of the spectral emission of Scorpius X-1 (hereafter Sco X-1), the brightest X-ray persistent source in the sky. Identified as the first X-ray extra-solar sources by \cite{giacconi_62}, Sco X-1 is a NS-LMXB system classified as Z-source \citep{Hasinger_89} in which the companion star is an M-type star with a mass of $\sim$ 0.4 M$_{\odot}$ \citep{steeghs_02}.
Sco X-1 was also the first X-ray binary found to exhibit radio emission \citep{Andrew_68} and, thanks to a monitoring campaign performed with the \textit{Very Long Baseline Array}, \cite{bradshaw_99} inferred a distance to the source of $2.8 \pm 0.3$ kpc\footnote{A more recent estimation of the distance can be inferred by the parallax measured by GAIA, resulting to be $2.1 \pm 0.1 $ kpc. Please, see the GAIA EDR3 catalogue at \url{https://gaia.ari.uni-heidelberg.de/tap.html}.}, while \cite{fomalont_01} determined an inclination angle $\theta = 46^{\circ} \pm 6^{\circ}$ of the system with respect to the line of sight.

Similar to all Z-sources, Sco X-1 exhibits QPOs along all the branches of the Z-track: observed for the first time by \cite{Middleditc_86}, we can distinguish, in general, between low ($<$10 Hz) and high ($>>$10 Hz) frequency QPOs and they were extensively studied.
The horizontal branch oscillations (HBOs) in Sco X-1 were observed for the first time by \cite{vanderklis_96}, with a peak of 45 Hz (and an harmonic near 90 Hz), then a \textit{twin} kHz HBOs were also detected in the range 800-1100 Hz, shifting simultaneously in frequency with constant peak-to-peak separation \citep{zhang_06,Yin_07,Yin_zhao_07}. The normal branch oscillations (NBOs) and the flaring branch oscillations (FBOs) were observed with peak frequencies in the range 4.5-7 Hz and 6-25 Hz, respectively, and seemed to be related to each other since the two peak frequencies converge when the source moves from NB to FB \citep[see e.g.][]{Casella_06,Yu_07}.

Because of the strong brightness of Sco X-1, which makes hard to collect high statistical data in the soft X-ray band without damaging the instruments, the study of the spectral emission of this source was mainly directed to the highest energies.
Using several HEXTE \citep[on-board \textit{Rossi X-ray Timing Explorer} satellite,][]{Rothschild_98} observations, \cite{damico_01} searched for the hard tail in the spectra of Sco X-1, modelling the data with a thermal bremsstrahlung model. The authors found that the addition of a power-law was necessary to model the data in 5 out of the 16 analysed observations.
\cite{barnard_03} used both HEXTE and PCA instruments to study the broadband spectrum of this source, fitting the data with a model composed of a blackbody from the NS and a cut-off power-law interpreted as a Comptonised emission from an extended accretion disc corona \citep[ADC,][]{white_82}, plus a broad Gaussian line. 
Whilst, \cite{bradshaw_03} applied a model composed of a blackbody emission plus a bulk motion Comptonisation and a broad Gaussian line to perform the analysis on PCA data in the range 2.5-18 keV.
\cite{disalvo_06} exploited the monitoring carried out by \textit{INTEGRAL} to follow the spectral evolution of the source along the Z-track. The authors analysed the data in the 20-200 keV energy range, collected by IBIS/ISGRI \citep{Ubertini_03,Goldwurm_03} during two entire revolutions of the satellite ($\sim$ 300 ks each one), using a thermal Comptonisation model and observing that the spectra were dominated above 30 keV by a power-law of photon index $\Gamma \sim 3$, with intensity slightly decreasing from the HB along the other branches, becoming not significant in the FB. The absence of a cut-off detection at the highest energy suggested the non-thermal origin of this hard component.
\cite{dai_07} analysed 43 spectra collected by \textit{RXTE} between 1997 and 2003, using a thermal and a hybrid Comptonisation model, plus a Gaussian and power-law component. The authors followed the spectral variation along the Z-pattern and obtained that, also in this case, the flux of the hard X-ray tail (with photon index between 1.9-3) was correlated with the position of the source in the CD and the contribution of this component to the total flux was anti-correlated to the mass accretion rate.
\cite{Church_12} exploited PCA+HEXTE observations to test the ADC model for Sco X-1 and Sco-like sources, finding that the behaviour of these ones in their NB is pretty similar to Cyg-like sources. The authors proposed also a general model for the Z-sources, in which the mass accretion rate does not increase monotonically along the Z-track but it rises from the soft apex to the hard apex, determining a constant high luminosity and high NS temperature, and consequently a high radiation pressure that causes the emission of relativistic radio-jets also in the FB.
\cite{titarchuk_14} followed the evolution of Sco X-1 between the HB and FB studying several \textit{RXTE} observations collected between 1996 and 2002. The authors fitted the 3-250 keV spectra with a model consisting of two Comptonised component with different seed photons temperature and a broad iron line, observing a stability in the value of the photon index during the HB and NB until a slight decreasing in the FB.
Finally, \cite{revnivtsev_14} studied 4 Ms of data collected by SPI \citep{Vedrenne_03} and IBIS instruments on-board \textit{INTEGRAL} and related simultaneous \textit{RXTE} observations, in order to have a coverage from 2 keV up to 10 MeV. The authors showed that the hard tail was well described by a power-law shape without cut-off up to 200-300 keV, proposing that it originates as a Compton up-scattering of soft seed photons on electrons with a non-thermal distribution.

\cite{homan_18} partially analysed the \textit{NuSTAR} observation of Sco X-1 reported here, in order to compare the results with those obtained in their study of the Z-source GX 5-1. Here, we extended the analysis of \textit{NuSTAR} data, studying the 3-60 keV spectra extracted for NB and FB with two different classes of models which lead to the same physical description of the source.

\section{Observations and data reduction}
The Nuclear Spectroscopic Telescope Array satellite \citep[\textit{NuSTAR},][]{harrison_13} observed Sco X-1 between 2014-10-08 06:46:07 UTC and 2014-10-08 16:16:07 UTC (ObsId 30001040002), for a total of $\sim$ 30 ks. The data were processed using the \textit{NuSTAR} Data Analysis Software (NuSTAR-DAS) v1.9.3 for both the data sets collected by the two focal plane modules, FPMA and FPMB. 

The source events were extracted from a circular region with radius 250'' and 200'' for FPMA and FPMB, respectively, centred on the source coordinates. While for the background events, we used a circular region with a radius of 120'' far away from the source for both the instruments. Because of the extremely high count rate of the source, the {\tt statusexpr} parameter in {\tt nupipeline} was modified to avoid artificially vetoing events from the noise filter.

The filtered events, the background-subtracted light curves, the spectra and the arf and rmf files were created using the {\tt nuproducts} tool; live-time, point-spread-function, exposure, and vignetting corrections were applied. 

\begin{figure}[!htbp]
    \centering
    \includegraphics[scale=0.58]{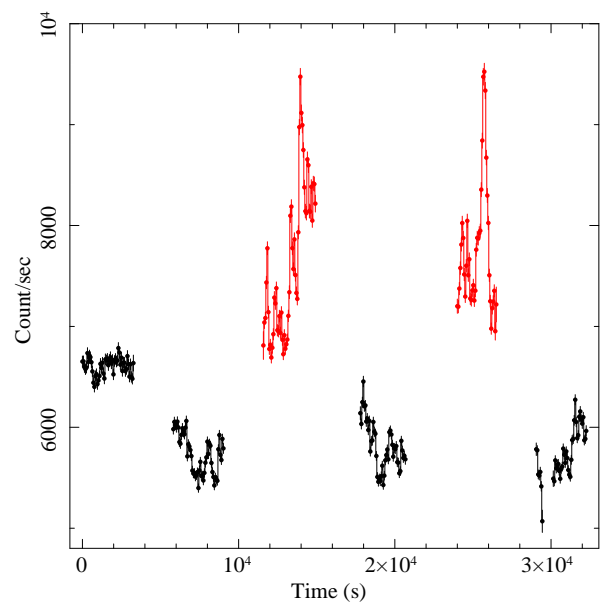}
    \caption{FPMA background-subtracted light curve of Sco X-1 in the 1.6-80 keV energy range, showing the flaring activity of the source (in red). The bin time is 64 s.}
    \label{fig:FPMA_lc}
\end{figure}

We observed the flaring activity of the source in both the FPMA and FPMB light-curve, with a count rate between 5000 c/s and 12500 c/s.
We show the 1.6-80 keV FPMA background-subtracted light curve in \autoref{fig:FPMA_lc}.
\begin{figure}[!htbp]
    \centering
    \includegraphics[scale=0.58]{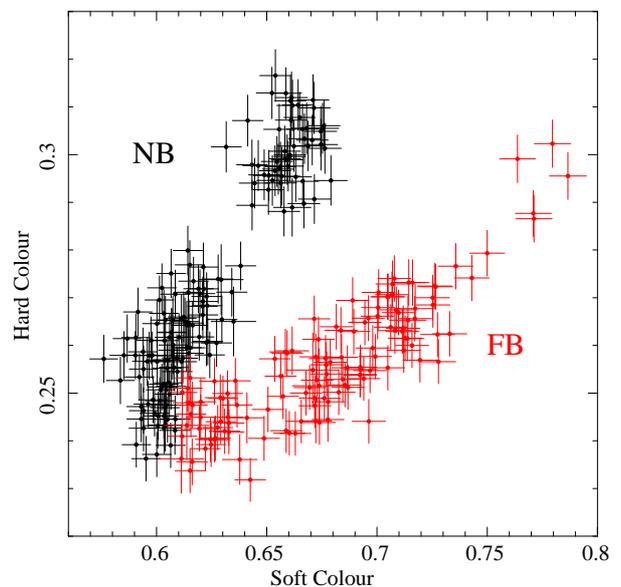}
    \caption{Colour-colour diagram of Sco X-1 from combined FPMA and FPMB data. The Soft Colour is the ratio between the count rate in the energy bands 6-10 keV and 3-6 keV, while the Hard Colour is the ratio between the count rate in the energy bands 10-20 keV and 6-10 keV.
    The bin time is 128 s. }
    \label{fig:FPMA_CCD}
\end{figure}
Furthermore, we build the CD of the source for the two instruments using the hardness ratio between the count-rate in the 6-10 keV and 3-6 keV energy bands and in the 10-20 keV and 6-10 keV energy bands to obtain the ``soft colour'' (SC) and ``hard colour'' (HC) light curves, respectively. 

Comparing the \textit{NuSTAR} CD with previous analysis reported in literature, we observe the same Z-track shape highlighted by CDs and HIDs obtained from \textit{RXTE} data (although the used energy ranges and the intensity are slightly different), in which only the lower normal branch and the flaring branch are visible and there are no evidence of the HB \citep[see e.g.][]{Church_12,Ding_21,Wang_21}. We combined the CD of the two instruments and we divided the two branches, performing a selection in HC and SC intervals in order to group the data. In particular, we selected the data between 0.58-0.62 SC and 0.22-0.25 HC, between 0.55-0.645 SC and 0.25-0.285 HC, and between 0.58-0.7 SC and 0.285-0.35 HC for NB (black points in \autoref{fig:FPMA_CCD}); for FB (red points in \autoref{fig:FPMA_CCD}) we selected the data in the following intervals: 0.62-0.66 SC and 0.22-0.25 HC, 0.66-0.74 SC and 0.225-0.228 HC and 0.74-0.82 SC and 0.25-0.33 HC. The grouped data are shown in \autoref{fig:FPMA_CCD} using different colours. 

Given the high count rate of the source, we explored the possibility of pile-up effects in our data. As reported by \cite{Grefenstette_16}, a possible pile-up effect can occur in two different situations: when two photons arrive in the same pixel and they are counted as one, or when two photons arrive in adjacent pixels and they are then combined during the reprocessing stage of the data.

For the first type of pile-up, \cite{Grefenstette_16} estimated for this \textit{NuSTAR} observation a pile-up fraction of 5.3$\times$10$^{-4}$ per pixel, that is a negligible contribution. For the second one, the same authors estimated a pile-up fraction of 8$\times$10$^{-4}$ per pixel, taking into account only the events with values of the 
 {\tt grades} parameter (that is a qualifier assigned to each event to identify the pattern of the detected photon on a $3\times3$  grid) between 21-24. As suggested by \cite{Grefenstette_16}, we extracted the spectrum of the source using {\tt grades=0-32} (total events), { \tt grades=0} (single-photon events only) and {\tt grades=1-32} (multiple-photons events only), but we did not observe a variation in the counts per energy channel related to a possible pile-up.

Using the standard {\tt nuproducts} pipeline, we obtained a total spectrum  with an exposure time around 800 s for both FPMA and FPMB, because of the dead-time correction factor due to the high count rate of the source. Finally, we extracted a NB spectrum of $\sim$500 s exposure time and a FB spectrum of $\sim$276 s exposure time, for both FPMA and FPMB.

\section{Data Analysis}
Since the spectral shape for the two instruments turn out to be compatible above 3 keV, they were fitted simultaneously in the 3-60 keV energy range, in order to avoid the predominant contribution of the background at the highest energies; the spectra, indeed, result background-dominated above 60 keV. All the spectra were grouped to have a minimum of 25 counts per energy bin. 

\begin{table*}[!htbp]
\centering
\begin{threeparttable}
\scriptsize
\caption{Best-fit results}
\begin{tabular}{lcccc|cccc}
\hline
\hline  
& & & & & \\
& \multicolumn{4}{c|}{NB} & \multicolumn{4}{c}{FB} \\
 & Model 2A$^{\star}$ & Model 2B$^{\diamond}$ & Model 3A$^{ \dagger}$ & Model 3B$^{ \ddagger}$ & 
 Model 2A$^{\star}$ & Model 2B$^{\diamond}$ & Model 3A$^{ \dagger}$ & Model 3B$^{ \ddagger}$ 
 	\\
 Component & & & & \\
 & & & & & \\
 {\sc TBabs} & & & & \\
  nH(10$^{22}$) & 0.3 (frozen)& 0.3 (frozen)& 0.3 (frozen)& 0.3 (frozen)& 0.3 (frozen)& 0.3 (frozen)& 0.3 (frozen)& 0.3 (frozen) \\ 
  & & & & & \\
  
  {\sc expabs} & & & & \\
  LowECut(keV) & & & 3 $kT$ & $kT_{\rm bb}$ &
    & & 3 $kT$ & $kT_{\rm bb}$\\ 
  & & & & & \\
{\sc powerlaw} & & & & \\
 PhoIndex & & & $3^{+2}_{-1}$ & $2 \pm 2$ &
 	& &$2 \pm 2$ & $2 \pm 2$
 \\
 norm & & & $>0.04$ & $<17$ &
  & & $< 32$ & $<123$
 \\
 F$_{\rm pow}$ ($\times$10$^{-9}$ erg cm$^2$s$^{-1}$)& & & 2.85& 1.06 & 
  & & 0.78 & 1.77
 \\ 
 & & & & & \\
 
{\sc Gaussian}& & & & \\
 E$_{\rm line_{\rm K\alpha}}$(keV) & 
 $6.62 \pm 0.03$ & $6.64 \pm 0.03$ & $6.61^{+0.03}_{-0.04}$ & $6.64 \pm 0.03$&
 $6.60 \pm 0.03$ & $6.63\pm 0.03$ & $6.60 \pm 0.03$ & $6.63 \pm 0.03$
\\
 $\sigma_{\rm K\alpha}$(keV) &
 $0.35^{+0.04}_{-0.05}$ & $0.33^{+0.04}_{-0.03}$ & $0.37^{+0.05}_{-0.04}$  & $0.35 \pm 0.04$ &
 $0.33 \pm 0.04$ & $0.28 \pm 0.04$ & $0.34^{+0.05}_{-0.04}$ & $0.29 \pm 0.04$
 \\
 norm & $0.12 \pm 0.01$ & $0.12 \pm 0.01$ & $0.13^{+0.02}_{-0.01}$ & $0.127 \pm 0.008$ &
 	 $0.16 \pm 0.02$ & $0.13 \pm 0.02$ & $0.162^{+0.018}_{-0.009}$ & $0.13 \pm 0.02$
 \\ 
 EqW$_{\rm K\alpha}$ (eV) & & &  $56^{+6}_{-7}$ & $54^{+7}_{-6}$ &
&  & $52^{+5}_{-6}$ & $41^{+6}_{-5}$
 \\ 
 & & & & & \\
 
{\sc Gaussian} & & & &\\
 E$_{\rm line_{\rm K\beta}}$ &
 $7.7 \pm 0.1$& $7.75 \pm 0.09$ & $7.8 \pm 0.1$ & $7.77 \pm 0.09$ & 
 $7.8 \pm 0.2$ & $7.8 \pm 0.3$ & $7.8 \pm 0.2$ & $7.9 \pm 0.2$ 
  \\
  
 norm &
 $0.022 \pm 0.006$ &  $0.026^{+0.007}_{-0.005}$ & $0.026^{+0.006}_{-0.003}$ & $0.031 \pm 0.004$ &
 	 $0.016 \pm 0.008$ & $0.010 \pm 0.005$ & $0.018 \pm 0.008$ & $0.013 \pm 008$
 \\
  EqW$_{\rm K\beta}$ (eV) &  & & $17 \pm 4$ & $21\pm 5$ &
 &  & $9^{+4}_{-5}$ & $6^{+5}_{-4}$
 \\ 
& & & & & \\
 
{\sc bbodyrad} & & & & \\
 $kT$ (keV) & $1.29 \pm 0.02$ & &  $1.26 \pm 0.02$ & &
    $1.51 \pm 0.01$ & & $1.50^{+0.02}_{-0.03}$ &
 \\

  R$_{\rm bb}$ (km) &$13 \pm 1$ & & $13 \pm 2$ & &
 	 $13 \pm 1$ & &$13 \pm 1$ &
 \\
 & & & & & \\
 
 {\sc diskbb} & & & & \\
 $kT_{\rm in}$ (keV) &
  & $0.46^{+0.02}_{-0.04}$ & & $0.43 \pm 0.03$ &
    & $0.62 \pm 0.03$ & & $0.60^{+0.04}_{-0.03}$  \\
 
 R$_{\rm disc}$ (km) & & $202^{+93}_{-40}$ & & $266^{+104}_{-83}$ & 
    & $98^{+22}_{-16}$& & $107 \pm 21$\\
 & & & & & \\
  
 F$_{\rm bb}$ (10$^{-7}$ erg cm$^2$s$^{-1}$) & & & 0.61 & 4.21 &
  &  & 1.19 & 2.76
  \\
  & & & & & \\
  
{\sc nthComp} & & & & \\
$\Gamma$ & 
$2.06 \pm 0.01$ &$2.45^{+0.02}_{-0.04}$ & $2.04^{+0.01}_{-0.02}$ & $2.40^{+0.03}_{-0.01}$ &
    $2.078^{+0.009}_{-0.018}$ & $2.87^{+0.06}_{-0.08}$ & $2.06^{+0.02}_{-0.01}$ & $2.77^{+0.09}_{-0.07}$
\\
$kT_{\rm e}$(keV) &
$2.96^{+0.02}_{-0.01}$ & $3.19^{+0.01}_{-0.04}$  & $2.88 \pm 0.04$ &$3.1^{+0.05}_{-0.03}$ &
    $2.99 \pm 0.04$ & $3.34^{+0.08}_{-0.09}$ & $2.94 \pm 0.06$ & $3.20^{+0.05}_{-0.10}$
\\
$kT_{\rm bb}$(keV) & $<0.33$ & $0.87^{+0.01}_{0.03}$ & $<0.34$ & $0.84 \pm 0.02$ &
	$< 0.38$ & $1.17^{+0.02}_{-0.03}$ & $<0.37$ & $1.15^{+0.03}_{-0.02}$
\\
norm & $99^{+4}_{-20}$ &$13.0^{+0.9}_{-0.4}$ & $93^{+6}_{-16}$ & $14.0^{+0.6}_{-0.7}$&
	 $107^{+7}_{-23}$ & $8.9^{+0.4}_{-0.3}$ & $109^{+3}_{-7}$ & $9.2^{+0.2}_{-0.8}$ 
\\
 & & & & & \\
	 
F$_{\rm Comp}$ (10$^{-7}$ erg cm$^2$s$^{-1}$) & & & 5.55 & 3.21 &
& & 6.8 & 3.67\\ 
 & & & & & \\
	 
$\tau$ & & & $9.8 \pm 0.2$& $7.58 \pm 0.09$ &
	& & $9.6 \pm 0.2$ & $6.2 \pm 0.3$
	\\ 
 & & & & & \\
	 	
F$_{\rm bol}$ (10$^{-7}$ erg cm$^2$s$^{-1}$) &
 & & 6.20 & 7.55 &
 & & 8.03 & 6.46
\\
	 
 \hline 
& & & & & \\
 $\chi^2/dof$ & 1395/1181 & 1339/1181 & 1336/1179 & 1307/1179 &
     1181/1099 & 1154/1099 & 1167/1097 & 1146/1097
 \\ 
 \hline \hline
\end{tabular}
\begin{tablenotes}
    \item[$^\star$] Model 2A:{\sf TBabs*(Gaussian+Gaussian+bbodyrad+nthComp)}
    \item[$^\diamond$]Model 2B:{\sf TBabs*(Gaussian+Gaussian+diskbb+nthComp)}
    \item[$^\dagger$]Model 3A: {\sf TBabs*(expabs*powerlaw+Gaussian+Gaussian+bbodyrad+nthComp)}
    \item[$^\ddagger$]Model 3B: {\sf TBabs*(expabs*powerlaw+Gaussian+Gaussian+diskbb+nthComp)}
    \item \textrm{The uncertainties are reported at 90\% confidence level. The spectral parameters are defined as in {\sc XSPEC}.
    \item \textrm{To estimate the value of the black body radius R$_{\rm bb}$ and the inner radius of the accretion disc R$_{\rm disc}$, we assumed a distance to the source of $2.8 \pm 0.3$ kpc (Bradshaw et al. 2003) and an inclination angle with respect to the line of sight $\theta=46^{\circ} \pm 6^{\circ}$ \citep{fomalont_01} .}
   \item \textrm{F$_{\rm pow}$, F$_{\rm bb}$ and F$_{\rm Comp}$ are the unabsorbed bolometric flux of the power-law, the thermal component and the Comptonised component in the 0.1-100 keV energy range, respectively. F$_{\rm bol}$ is the total unabsorbed bolometric flux in the 0.1-100 keV energy range.} 
    }
\end{tablenotes}
    \label{tab:fits}
    \end{threeparttable}
\end{table*}

We used {\sc XSPEC} v12.10.1q to perform the spectral fit; we set the abundances and the photoelectric absorption cross-sections to the values found by \cite{wilms_00}  and by \cite{verner_96}, respectively, for all the spectral models discussed in the following. We applied the same models to the NB and the FB spectra.
\begin{figure*}[!htbp]
    \centering
    \includegraphics[scale=0.5]{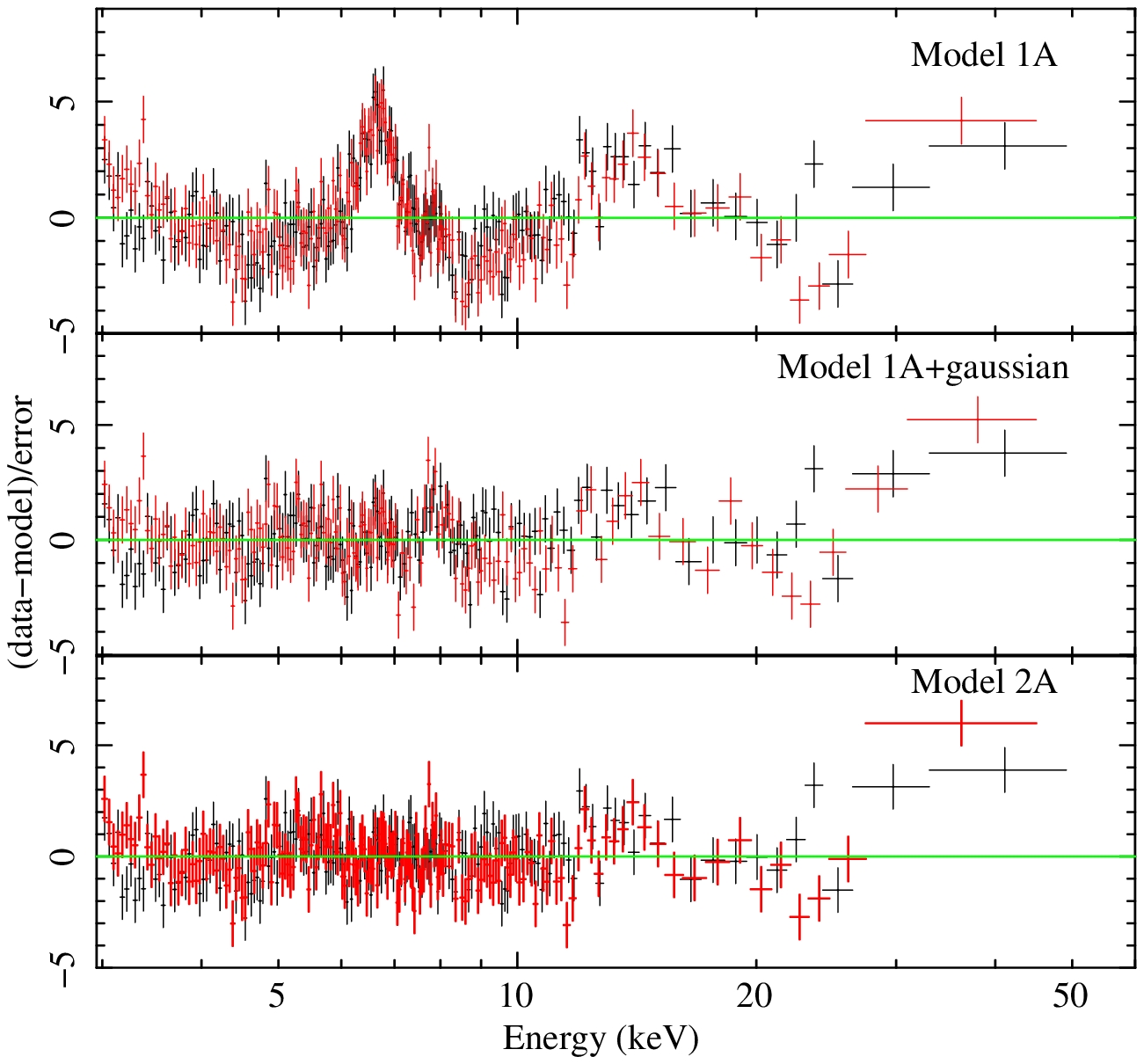}
    \hspace{0.5 cm}
    \includegraphics[scale=0.5]{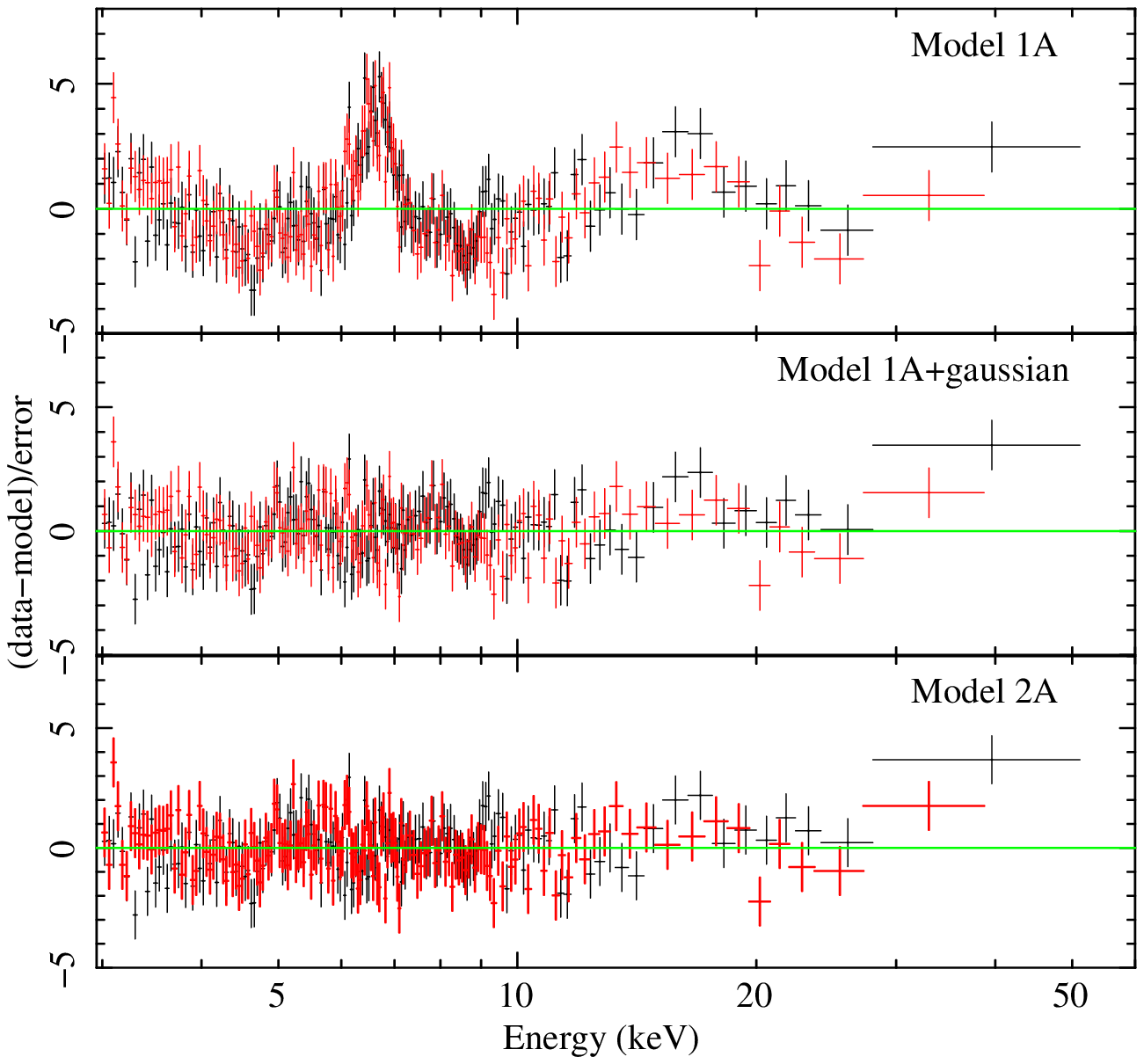}
    \caption{On the left: comparison between residuals obtained adopting Model 1A (top panel), Model 1A plus a Gaussian component (middle panel) and Model 2A (bottom panel) for NB.
    On the right: same comparison for FB.
    The FPMA and FPMB data are showed in black and red colour, respectively.
    The residuals are graphically re-binned in order to have at least 100$\sigma$ per bin.}
    \label{fig:residuals_1A}
\end{figure*}
\subsection{Model 1A}
Initially, we fitted the continuum direct emission with a model composed of a blackbody component ({\tt bbodyrad} in {\sc XSPEC}), which mimics a saturated Comptonisation associated with a boundary layer, plus a thermal Comptonisation \citep[{\tt nthComp},][]{zdziarski_96,zychi_99}, with the {\tt inp\_type} parameter set to 1, indicating that the seed photons have a disc-blackbody distribution.
To take into account the photoelectric absorption by neutral matter in the ISM, we used the T{\"u}bingen-Boulder model ({\tt TBabs} component), keeping the value of the equivalent hydrogen column associated with the interstellar matter fixed to 0.3$\times$10$^{22}$ cm$^{-2}$ \citep{Ding_21,Church_12,dai_07,Christian_97} due to the lack of coverage below 3 keV.
From this model, called Model 1A, we obtained a $\chi^2/$d.o.f. of 2149/1186 and 1772/1104 for NB and FB, respectively, but we observed large residuals around 6 keV, in the Fe-K region (top left and top right panels in \autoref{fig:residuals_1A}).

We added, then, a Gaussian component to the model, leaving all the line parameters free to vary. We obtained a significant improvement of the fit, with a $\chi^2/$d.o.f. of 1428/1183 and 1190/1101 for NB and FB, respectively, and a significance of the line component (estimated as the ratio between the intensity of the line and the associated error calculated at 68\% c.l.) of $13\sigma$ for NB spectrum and $15\sigma$ for FB spectrum.
The emission line showed a centroid energy around 6.6 keV and a line width $\sigma \sim 0.3$ keV.

\subsection{Model 2A}
Looking at the residuals (central panels in \autoref{fig:residuals_1A}), we observe still some features in the Fe-K region, especially in the NB spectrum. Then we added a second Gaussian component, keeping linked the width to the $\sigma$ parameter of the first one, under the hypothesis that the two lines are associated to the emission from the same ion in the same region of the accretion flow, and so that 
their dispersion velocity is not dependent from the atomic weight of the element and they are affected by the same broadening effects.
From this Model 2A we obtained $\chi^2/$d.o.f.$=1395/1181$ with a significance of the new line component of $7.3\sigma$ for NB and $\chi^2/$d.o.f.$=1181/1099$ with a 3$\sigma$ of significance for FB, implying that the detection of this feature is barely significant in the flaring branch spectrum. The centroid energy of this line is around 7.8 keV, suggesting an emission related to the K$\beta$ transition of the \ion{Fe}{xxv} ion.
The best-fit results and residuals are shown in \autoref{tab:fits} (second and sixth columns for NB and FB, respectively) and \autoref{fig:residuals_1A} (bottom panels).

\begin{figure*}[!htbp]
    \centering
    \includegraphics[scale=0.5]{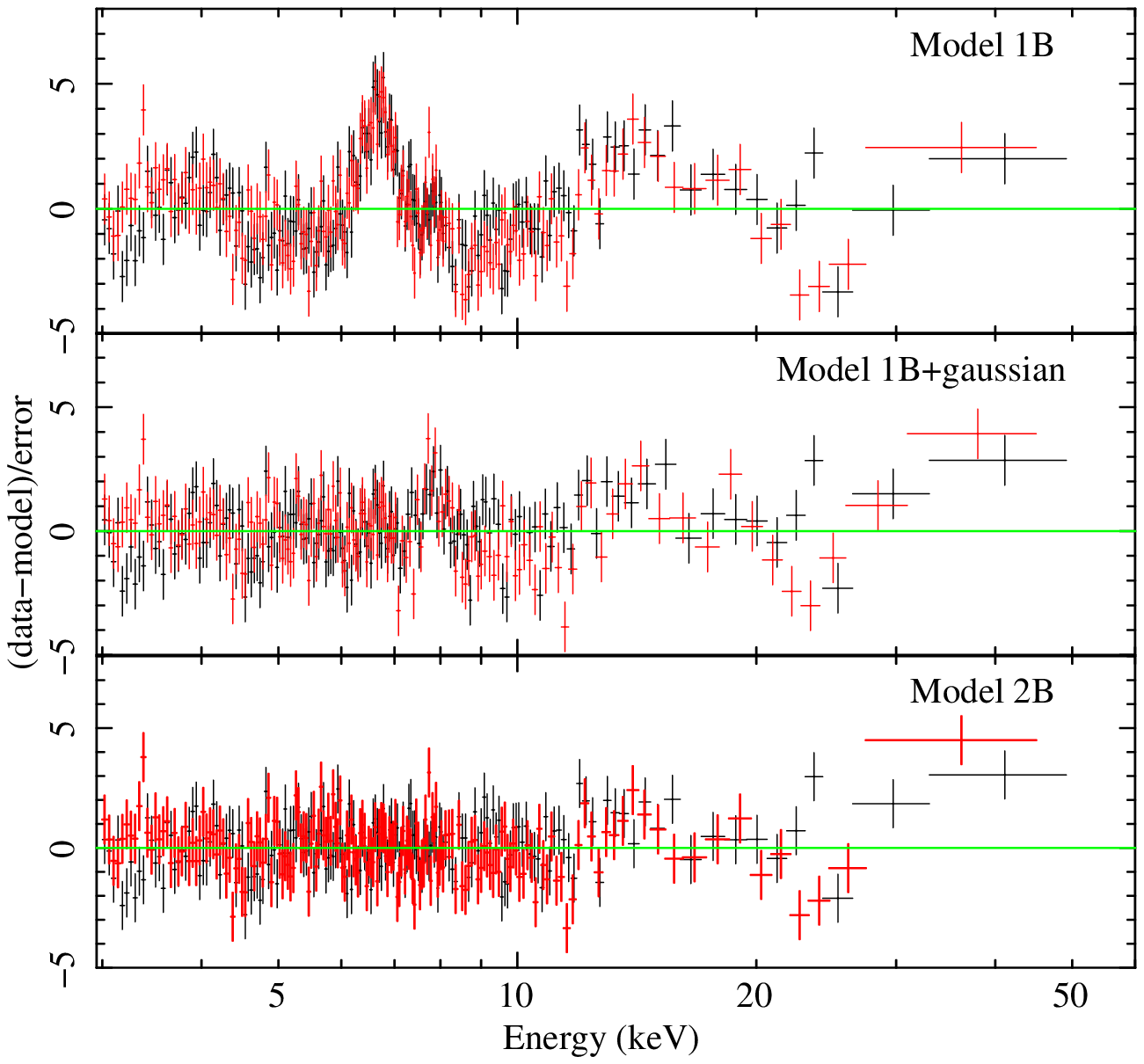}
    \hspace{0.5 cm}
    \includegraphics[scale=0.5]{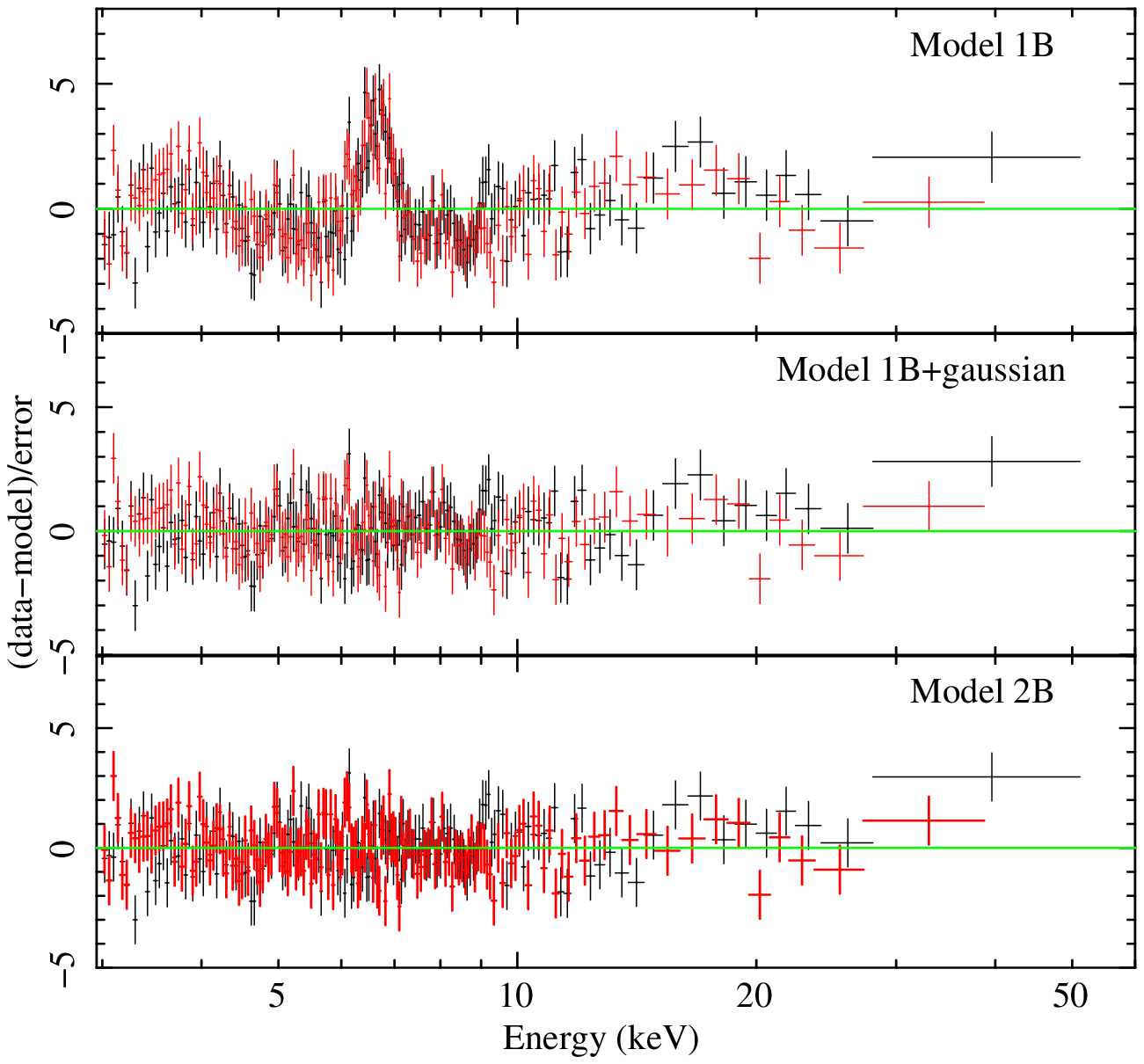}
    \caption{On the left: comparison between residuals obtained adopting Model 1B (top panel), Model 1B plus a Gaussian component (middle panel) and Model 2B (bottom panel) for NB.
    On the right: same comparison for FB.
    The FPMA and FPMB data are showed in black and red colour, respectively.
    The residuals are graphically re-binned in order to have at least 100$\sigma$ per bin.}
    \label{fig:residuals_1B}
\end{figure*}
\subsection{Model 3A}
Since we still observed large residuals above 30 keV, we added to Model 2A a {\tt powerlaw} component to fit this hard excess, leaving the photon index and the normalisation free to vary.
We obtained a $\chi^2/$d.o.f. of 1285/1179 and 1145/1097 for NB and FB, respectively, corresponding to a $\Delta \chi^2$ with respect to Model 2A of 110 and 36 (with 2 d.o.f. gap), which represent an improvement of the fits. Furthermore, we obtained an F-test probability of chance improvement of 8.8$\times$10$^{-22}$ for NB and of 4.6$\times$10$^{-8}$ for FB, corresponding to a significance larger than $6\sigma$ and about $5.7\sigma$, respectively, suggesting that the power-law component is required in both the cases.
We added also a low-energy exponential roll-off to the model, using the {\tt expabs} component, in order to mimic a cutoff at the seed photon temperature, keeping fixed the low energy cut-off to a value equal to 3 times the blackbody peak temperature $kT$; we called this one Model 3A. We obtained a $\chi^2/$d.o.f. of 1336/1179 with an F-test probability of chance improvement of 9$\times$10$^{-12}$ (significance $>6\sigma$) for NB  and $\chi^2/$d.o.f.=1167/1097 with an F-test of 0.0018 (significance $\sim3\sigma$) for FB. Also in this case, the statistical weight of the power-law decreases in the flaring branch spectrum. The significance of this component was confirmed also using Monte-Carlo simulations (please, see Appendix \ref{appendix}).

\begin{figure*}[!htbp]
    \centering
    \includegraphics[scale=0.5]{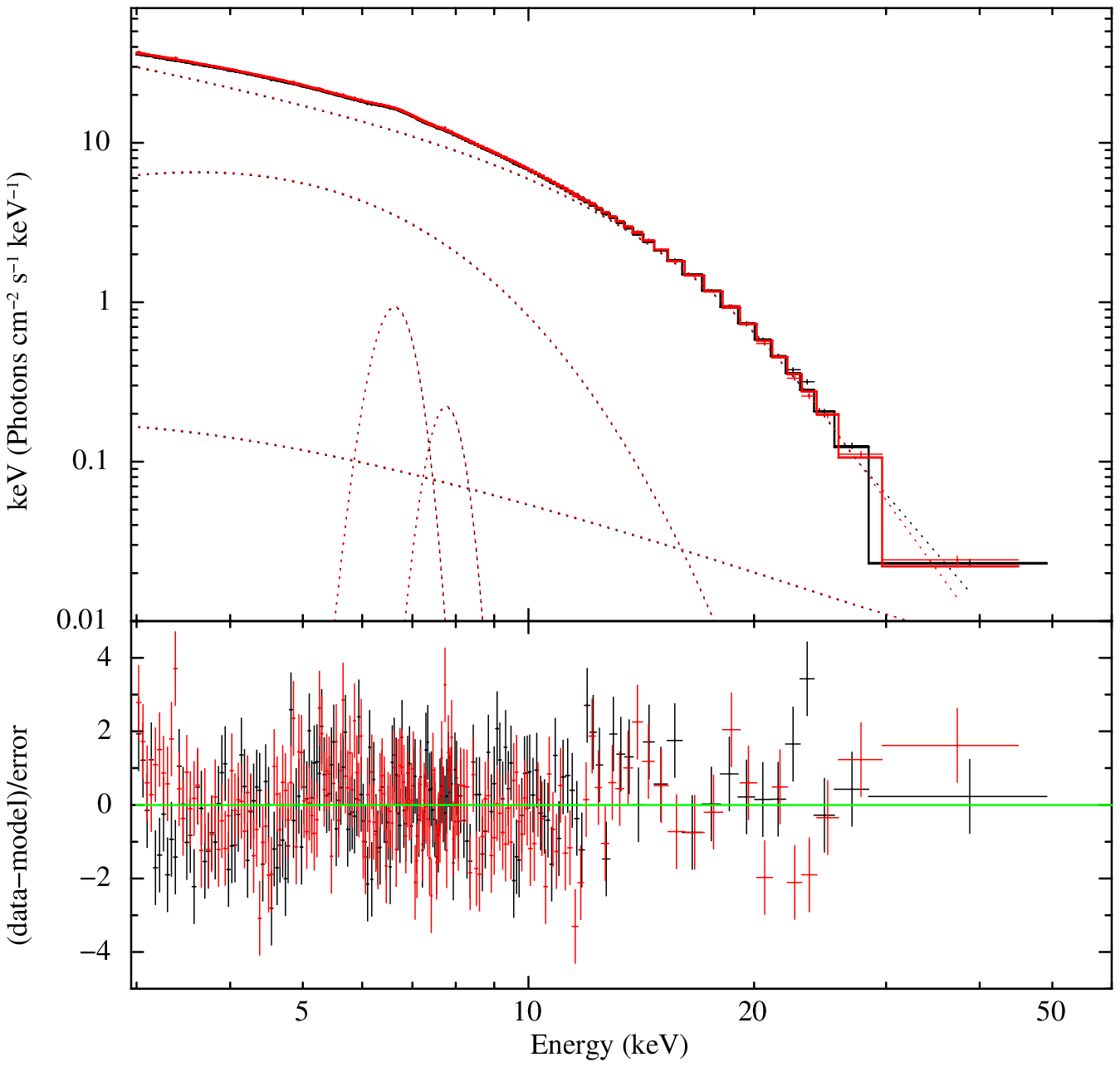}
    \hspace{0.5 cm}
    \includegraphics[scale=0.5]{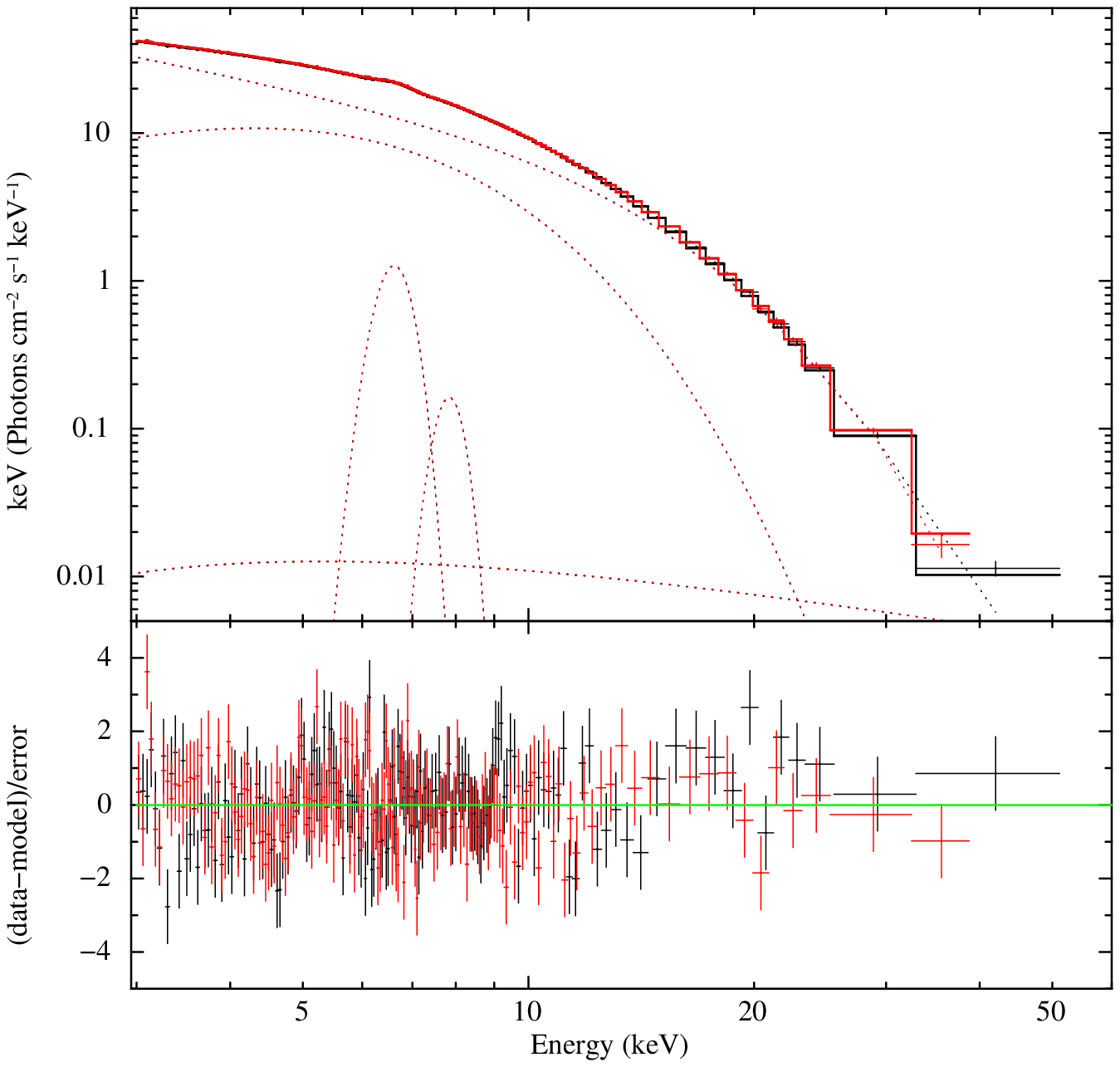}
    \caption{The unfolded spectrum and corresponding residuals obtained using Model 3A for NB (on the left) and FB (on the right).
    The FPMA and FPMB data are shown in black and red colour, respectively.
    The spectra and the residuals are graphically re-binned in order to have at least 100$\sigma$ per bin.}
    \label{fig:Model_4A}
\end{figure*}

For the NB spectrum, we observe a total unabsorbed flux $F_{\rm bol}= 6.20 \times 10^{-7}$ erg cm$^{-2}$ s$^{-1}$ in the 0.1-100 keV energy range, corresponding to a luminosity of $5.79 \times 10^{38}$ erg s$^{-1}$ for a distance to the source of $2.8 \pm 0.3$ kpc \citep{bradshaw_03}; the 0.1-100 keV bolometric flux associated to the power-law component is $F_{\rm pow}=2.85 \times 10^{-9}$ erg cm$^{-2}$ s$^{-1}$. While for the FB spectrum, the total unabsorbed flux in the 0.1-100 keV energy range is $F_{\rm bol}= 8.0 \times 10^{-7}$ erg cm$^{-2}$ s$^{-1}$ , corresponding to a luminosity of $7.5 \times 10^{38}$ erg s$^{-1}$, and the bolometric flux associated to the power-law component is $F_{\rm pow}=0.78 \times 10^{-9}$ erg cm$^{-2}$ s$^{-1}$.
The best-fit values of the parameters are shown in the fourth and eighth column of \autoref{tab:fits}; the unfolded spectrum and the corresponding residuals are shown in \autoref{fig:Model_4A}.

\subsection{B-Models}
As alternative description, we substituted the blackbody component with a multi-colour disc-blackbody \citep[{\tt diskbb} in {\sc XSPEC}, ][]{mitsuda_84,makishima_86}, varying accordingly the {\tt inp\_type} value of {\tt nthComp} to 0, indicating that the seed photons have a blackbody incoming spectrum; we called this Model 1B. Following the same steps described above, we added two Gaussian lines, obtaining our Model 2B:{\sf TBabs*(Gaussian+Gaussian+diskbb+nthComp)}. The K$\alpha$ line has a significance of 12$\sigma$ in both the branch spectra, while the K$\beta$ has a significance of 8.7$\sigma$ in NB and only 2$\sigma$ in FB; the values of the lines parameters are compatible with those obtained from Model 2A. The best-fit results are shown in \autoref{tab:fits} (third and seventh columns for NB and FB, respectively); the residuals for each model step are presented in \autoref{fig:residuals_1B}.
\begin{figure*}[!htbp]
    \centering
    \includegraphics[scale=0.5]{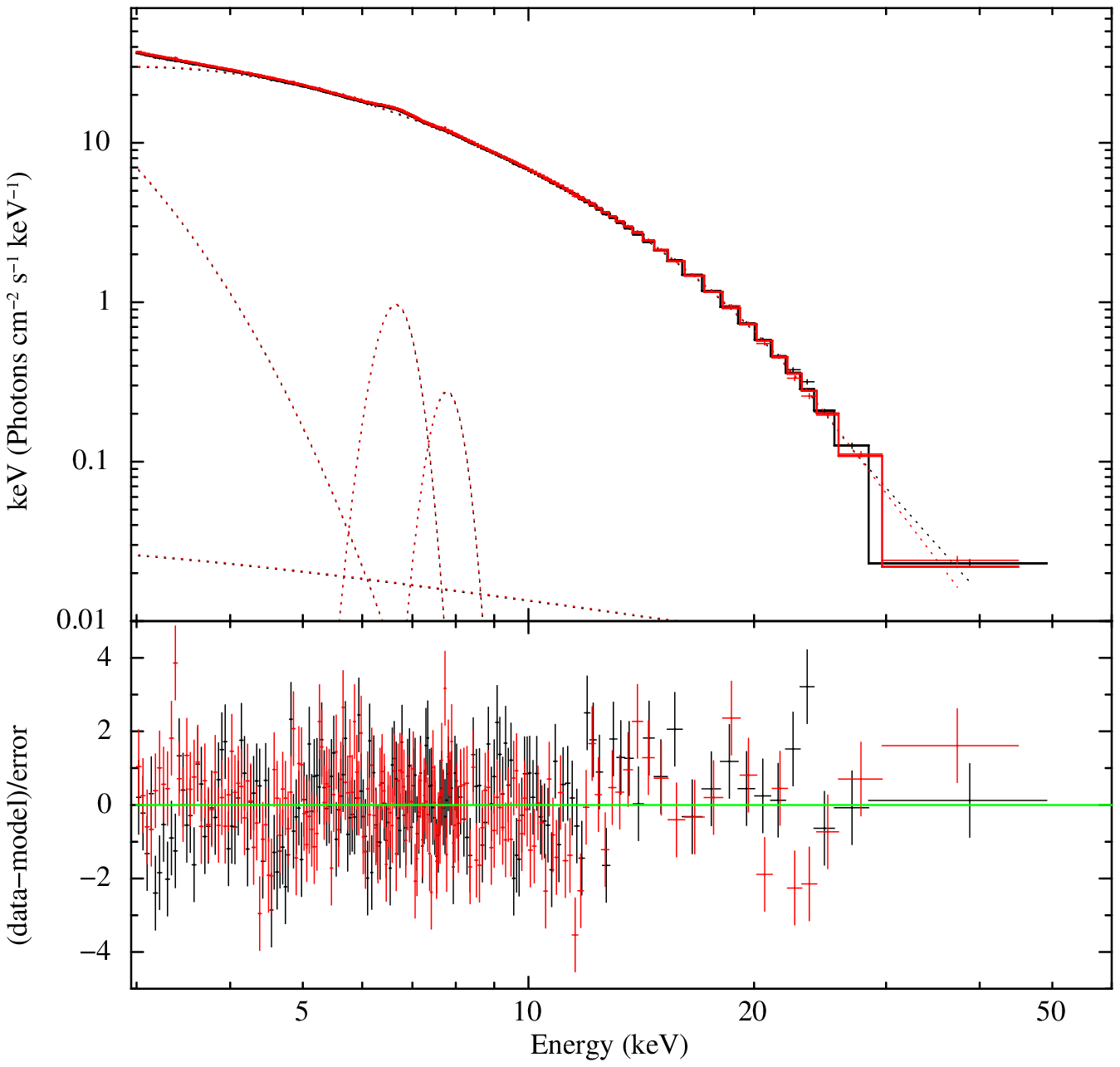}
    \hspace{0.5 cm}
    \includegraphics[scale=0.5]{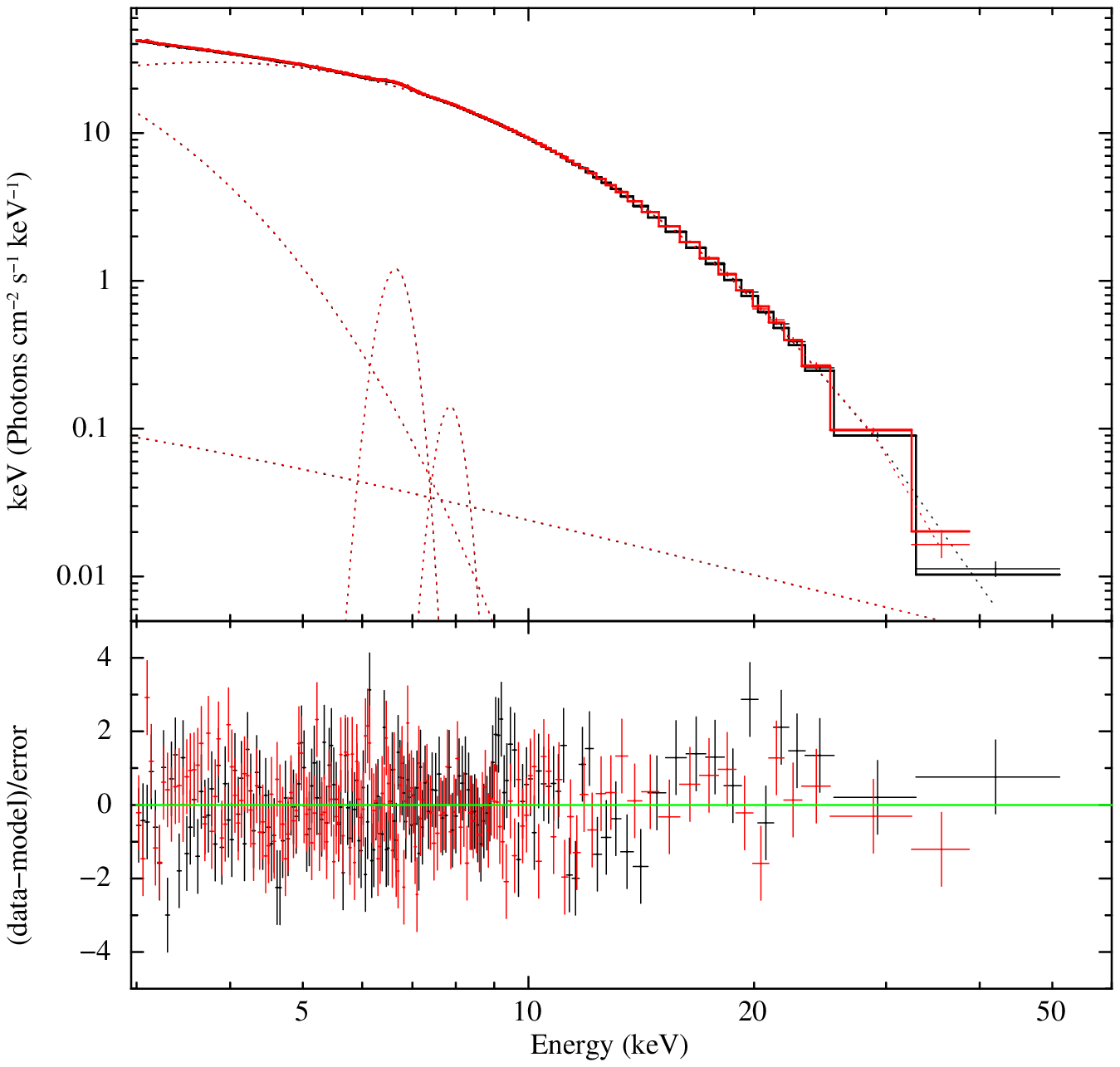}
    \caption{The unfolded spectrum and corresponding residuals obtained using Model 3B for NB (on the left) and FB (on the right).
    The FPMA and FPMB data are shown in black and red colour, respectively.
    The spectra and the residuals are graphically re-binned in order to have at least 100$\sigma$ per bin.}
    \label{fig:Model_4B}
\end{figure*}

Then, we added a power-law to the model, obtaining $\chi^2/$d.o.f.=1307/1179 for NB with an F-test probability of chance improvement of 6.7$\times$10$^{-7}$, corresponding to a significance of $\sim5\sigma$, which suggests that the addition of this component is statistically significant. While for FB we obtained $\chi^2/$d.o.f.=1146/1097 with an F-test of 0.0156 (significance $<2.8\sigma$), suggesting that the presence of the power-law is not significant.

The addition of {\tt expabs} is not required by the data in this case. However, in order to perform a self-consistent analysis, we decided to take into account this component also in this model, called Model 3B, keeping fixed the low energy cut-off to the seed photon temperature $kT_{\rm bb}$.

Also in this case, we test the significance or the power-law component using Monte-Carlo simulations (please, see Appendix \ref{appendix}).
The total unabsorbed flux observed in the 0.1-100 keV energy range is $F_{\rm bol}= 7.55 \times 10^{-7}$ erg cm$^{-2}$ s$^{-1}$ , corresponding to a luminosity of $7.1 \times 10^{38}$ erg s$^{-1}$, in NB and $F_{\rm bol}= 6.46 \times 10^{-7}$ erg cm$^{-2}$ s$^{-1}$ in FB, corresponding to a luminosity of $6.1 \times 10^{38}$ erg s$^{-1}$. The 0.1-100 keV bolometric flux associated to the power-law component is $F_{\rm pow}= 1.06 \times 10^{-9}$ erg cm$^{-2}$ s$^{-1}$ and $F_{\rm pow}= 1.77 \times 10^{-9}$ erg cm$^{-2}$ s$^{-1}$ in NB and FB, respectively.
The best-fit values of the parameters are shown in the fifth and ninth column of \autoref{tab:fits}; the unfolded spectrum and the corresponding residuals are shown in \autoref{fig:Model_4B}.

\subsection{Model 4}
Moving forward, we explored the hypothesis that the iron emission lines originates from a smeared reflection, thus we replaced the Gaussian components in Model 3B with a {\tt diskline} \citep{fabian_89}. We kept the outer radius of the reflection region $R_{\rm out}$ and the inclination angle $\theta$ of the binary system fixed to the value of 3000 gravitational radii ($Rg= G M/ c^2$) and 46$^{\circ}$ \citep{fomalont_01}, respectively, while the inner radius $R_{\rm in}$, the power-law dependence of emissivity {\tt  Betor} and the energy of the emission line were left free to vary.
In order to lead the fit to convergence, we kept the photon index of the power-law component fixed to the value obtained from Model 3B for both NB and FB spectrum.
We called this Model 4. We obtained a $\chi^2/$d.o.f. of 1374/1181 and 1155/1099 for NB and FB, respectively. 
The line energy and best-fit values of the continuum parameters are compatible at 90\% c.l. with the results obtained from Model 3B for both NB and FB. On the other hand, the inner radius $R_{\rm in}$ is not well constrained: we obtained only un upper limit of 14 Rg in NB and $R_{\rm in}=64^{+70}_{-53}$ Rg in FB.
The emissivity parameter is $-2.14^{+0.08}_{-0.13}$ for NB spectrum,  and {\tt  Betor}$=-2.5^{+0.5}_{-1.0}$ in FB, accordingly with the value expected for a LMXB to be between -2 (in the case of a dominating central illuminating flux) and -3, which describes approximately the intrinsic emissivity of a disc \citep[see e.g.][]{Dauser_13, Ponti_2018,DiSalvo_19,Mazzola_19, Marino_19,iaria_19,iaria_20}.

Slightly large residuals are observed around 7.8 keV, especially for the NB spectrum (see central panels in \autoref{fig:residuals_4_5}). Furthermore, we obtained a $\Delta \chi^2$ of 67 and 9 (2 d.o.f. apart) with respect to Model 3B for NB and FB spectrum, respectively. Thus, we do not get an improvement of the fit. 

\subsection{Model 5}
Then, we replaced the {\tt diskline} component with a self-consistent reflection model \citep[{\tt rfxconv},][]{kolhemainen_11}, in order to fit the emission line in the Fe-K region and take also into account the reflection continuum. We kept the iron abundance fixed to the Solar one, the redshift parameter $z$ fixed to 0 and the cosine of the inclination angle fixed to 0.6947, considering again $\theta=46^{\circ}$  \citep{fomalont_01}, while we left the ionisation parameter $\log \xi$ of the reflecting matter in the accretion disc and the relative reflection normalisation rel\_refl free to vary. The incident emission onto the accretion disc is provided by the Comptonisation component. 
We call this Model 5, the best-fit results are shown in \autoref{tab:reflection}.
We obtained a $\chi^2/$d.o.f. of 1316/1183 with a $\Delta \chi^2$ of 9 with respect to Model 4B for NB spectrum; while $\chi^2/$d.o.f.=1145/1101 and $\Delta \chi^2$=3 for FB spectrum, suggesting that the reflection model does not provide an improvement of the fit.
\begin{table}[!htbp]
    \centering
    \begin{threeparttable}
    \caption{Best-fit results from Model 5: { TBabs*(expabs*powerlaw+diskbb+rfxconv*nthComp)}}
    \small
 \begin{tabular}{lcc}
\hline
\hline
& NB & FB \\
Component & & \\
{\sc TBabs} \\ 
nH(10$^{22}$) & 0.3 (frozen) & 0.3 (frozen)\\ \\

{\sc powerlaw} \\
PhoIndex & 2 (frozen) &  2 (frozen)\\
 norm & $0.086^{+0.022}_{-0.022}$ & $0.5 \pm 0.3$ \\ \\
 
{\sc rfxconv} \\
rel\_refl & $0.20 \pm 0.03$ & $0.14 \pm 0.02$ \\
 log$\xi$ & $2.37^{+0.03}_{-0.02}$ & $2.48^{+0.09}_{-0.05}$ \\ \\
 
 {\sc diskbb} \\
 Tin(keV) & $0.55 \pm 0.03$ & $0.64 \pm 0.03$\\
R$_{\rm disc}$ (km) & $122 \pm 27$ & $89 \pm 15$ \\ \\
 
{\sc nthComp} \\
Gamma & $2.49 \pm 0.03$ & $2.90^{+0.09}_{-0.07}$\\
$kT_{e}$(keV) & $3.14^{+0.04}_{-0.05}$ & $3.29^{+0.11}_{-0.08}$ \\
$kT_{bb}$(keV) & $0.92 \pm 0.02$ & $1.19^{+0.03}_{-0.02}$ \\
norm & $10.9^{+0.5}_{-0.6}$ & $8.3^{+0.3}_{-0.4}$\\ \\

$\tau$ & $7.2 \pm 0.2$ & $6.0 \pm 0.3$ \\ \\
R$_{\rm seed}$ (km) & $39 \pm 4$ & $26 \pm 3$ \\
\hline
 $\chi^2/dof$ & 1316/1183 &  1149/1101 \\
\hline
\hline
\end{tabular}
\begin{tablenotes}
\scriptsize{
  \item  The parameters Fe\_abun (iron abundance), $z$ (redshift) and CosIncl (cosine of the inclination angle) of {\sc rfxconv}  were kept fixed to the values of 1, 0 and 0.6947, respectively.
 \item The uncertainties were calculated at 90\% c.l. The spectral parameters are defined as in {\sc XSPEC}.
    \item \textrm{To estimate the value of the inner radius of the accretion disc R$_{\rm disc}$, we made the same assumptions reported for the Model 3B.}
   \item The optical depth $\tau$ of the electron corona was estimated using the relation provided by \cite{zdziarski_96}, while for the seed-photon emitting radius R$_{seed}$ we apply the form by \cite{zand_99}.} 
\end{tablenotes}
    \label{tab:reflection}
     \end{threeparttable}
\end{table}

\begin{figure*}[!htbp]
    \centering
    \includegraphics[scale=0.5]{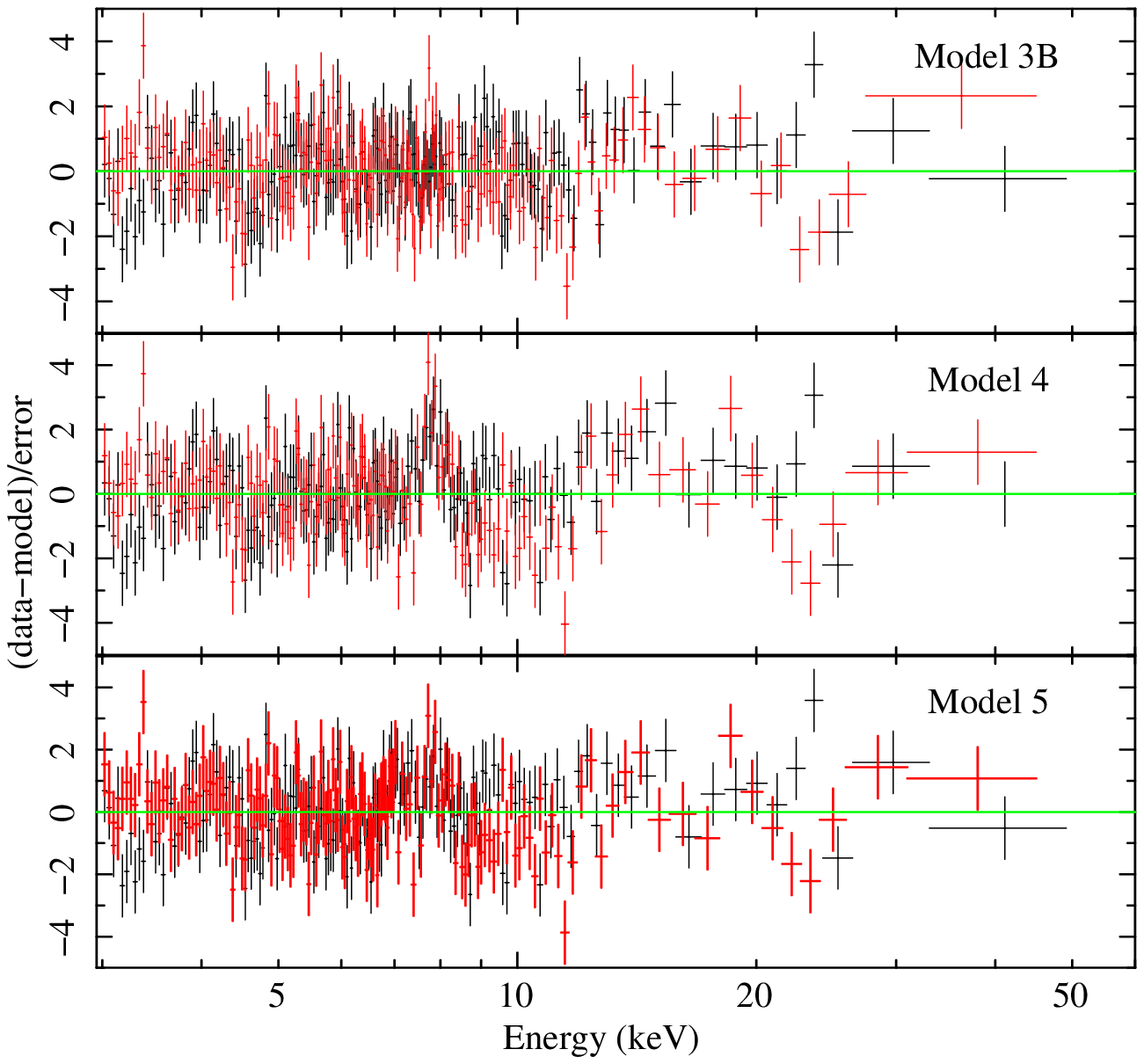}
    \hspace{0.5 cm}
    \includegraphics[scale=0.5]{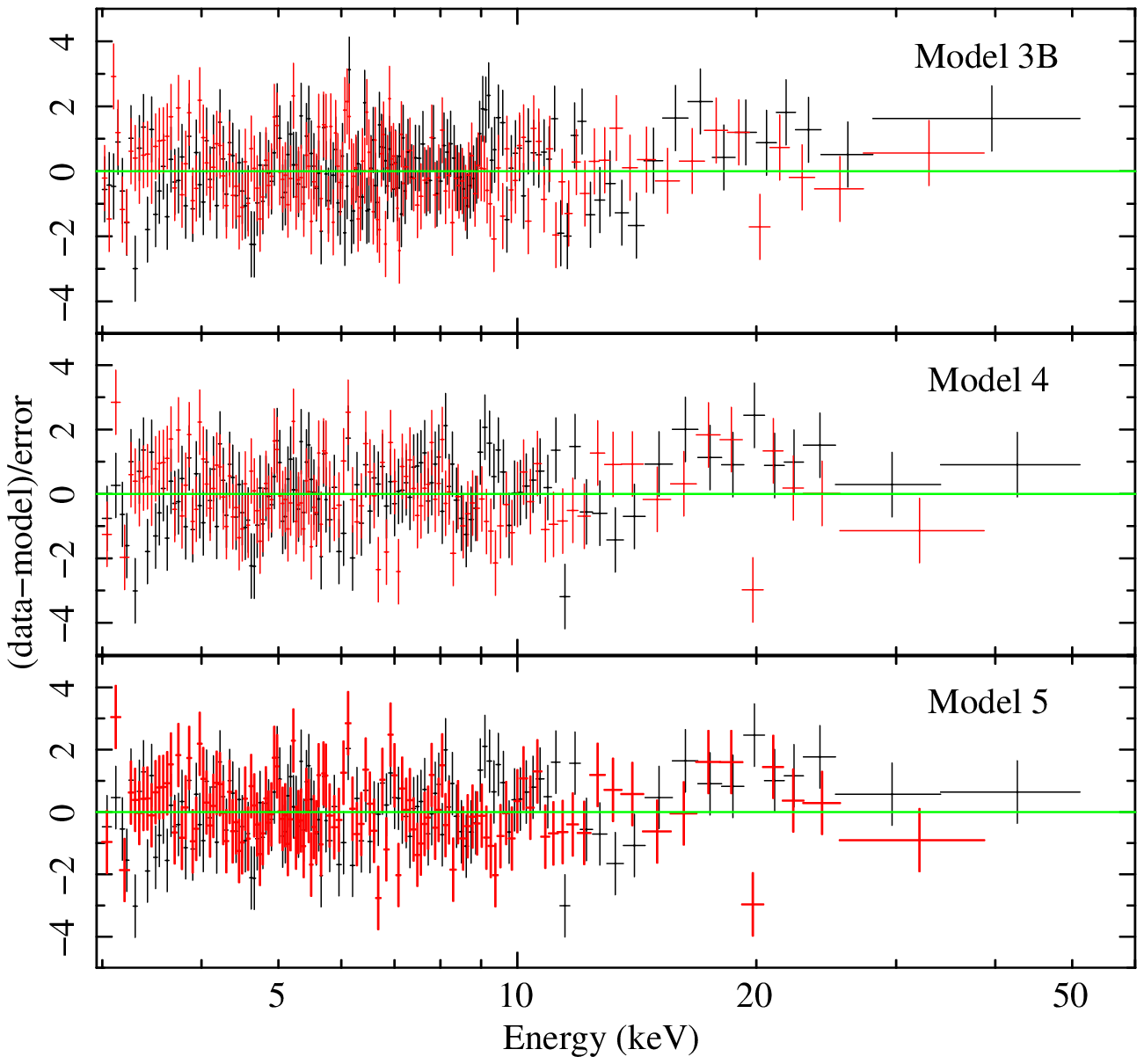}
    \caption{On the left: comparison between residuals obtained adopting Model 3B (top panel), Model 4 (middle panel) and Model 5 (bottom panel) for NB.
    On the right: same comparison for FB.
    The FPMA and FPMB data are shown in black and red colour, respectively.
    The residuals are graphically re-binned in order to have at least 120$\sigma$ per bin.}
    \label{fig:residuals_4_5}
\end{figure*}
To take into account the possible relativistic smearing effects in the inner region of the accretion disc, we added to Model 6 a multiplicative {\tt rdblur} component, using the same values of the parameters described for the {\tt diskline} in Model 4. Since we did not obtain any constrains on the inner radius R$_{\rm in}$ at which the reflection component originates, in the hypothesis that the reflection occurs in a region of the disc where the relativistic effects are not predominant, we arbitrary decided to tie the inner reflection radius to the inner radius of the accretion disc, related to the normalisation of the disc-blackbody component by the relation ${norm}= R_{\rm disc}^2/D_{10}^2 \ \cos \theta$, where $D_{10}$ is the distance to the source in unit of 10 kpc and $\theta$ is the inclination angle of the source. Under this assumption, we found a constrain on the emissivity factor {\tt  Betor} of $-2.0^{+0.5}_{-0.34}$ in NB and $> -1.7$ in FB.
 This model provides a $\chi^2/$d.o.f. of 1305/1182 with an F-test probability of chance improvement of 0.0005, corresponding to a significance $\sim 3\sigma$ for NB spectrum and $\chi^2/$d.o.f.=1146/1100 with an F-test value of 0.0698 (significance $<2 \sigma$) for FB spectrum. These results suggest that the addition of {\tt rdblur} component is not significant, then we can suppose that the iron emission line is not affected by relativistically smearing but only by Compton broadening.

Despite the best-fit values of the parameters are compatible at 90\% c.l. with those obtained from Model 3B for both the branches, from Model 5 we still observe slightly larger residuals around 7.8 keV, especially in NB spectrum (see the comparison in \autoref{fig:residuals_4_5}). For this reason, together with the evidence that the reflection does not produce an improvement for the fit, we consider the two Gaussian models (i.e. Model 3A and 3B) as our best-fit models.
Finally, we can infer that the Fe-K region of the spectrum is well-described by the presence of two Compton broadened iron emission line, related to the K$\alpha$ and K$\beta$ transition of the \ion{Fe}{xxv} ions in the accretion disc.

In the end, we tested also the model used by \cite{homan_18}, composed of two thermal component, a power-law and a Gaussian component to fit the iron emission line at 6.6 keV, plus a {\tt TBabs} component to take into account the absorption due to the interstellar medium. For the NB spectrum, we inferred a $\chi^2/$d.o.f. of 1380/1184, reproducing the same results obtained by \cite{homan_18}, but we did not attain an improvement of the fit with respect to Model 3A and 3B. For the FB spectrum, not analysed by \cite{homan_18}, we did not obtain a good fit using this model; we inferred a photon index of the power-law around 7 and large residuals remain in the Fe-K region of the spectrum and above 20 keV. 
The two blackbody components try to fit the residuals related to the K$\beta$ emission line as a soft excess, not expected from the source, and the power-law component alone is not sufficient to model both the Comptonised emission and the hard excess observed above 20 keV.

\section{Discussion}
We analysed the first \textit{NuSTAR} observation of the source Sco X-1, collected in October 2014. From the obtained CD, we pointed out that the observation covers the lower normal branch and the flaring branch of its Z-track; thus, we extracted an almost complete spectrum for NB and a partial spectrum for FB. For both the branches, we find that the 3-60 keV spectrum is well fitted by a model composed of a thermal emission, a thermal Comptonisation and two Gaussian components corresponding to the Fe K$\alpha$ and Fe K$\beta$ emission line of the \ion{Fe}{xxv} ion; a hard tail was detected above 20 keV and modelled by a power-law limited at low energies with a photon index value of 2. We adopted two different description for the thermal emission: a blackbody component (Model 3A) and a multi-colour disc-blackbody (Model 3B).

From model 3A, assuming a distance to the source of 2.8 kpc, we inferred a blackbody radius $R_{\rm bb} = 13 \pm 2$ km and $13 \pm 1$ km for NB and FB, respectively, according with the hypothesis that this emission generates in the innermost region of the system,  generally identified with the NS surface and the so-called boundary layer.
Furthermore, we estimated the radius of the seed-photon emitting region R$_{\rm seed}$ using the relation $R_{\rm seed}= 3 \times 10^4 d [F_{\rm Compt} /(1+y)]^{1/2} (kT_{\rm seed})^{-2}$ \citep{zand_99}, where $d$ is the distance to the source in units of kpc, $y= 4kT_e\ max[\tau,\tau^2]/(m_ec^2)$ is the Compton parameter, $kT_{\rm seed}$ is the seed-photon temperature in units of keV, and $F_{\rm Compt}$ is the bolometric flux of the Comptonisation component. Using the best-fit values (fourth and seventh column in \autoref{tab:fits}), we find $R_{\rm seed}=303 \pm 2$ km and $286 \pm 30$ km for NB and FB spectrum, respectively. In this case, we are assuming the seed photons come from an equivalent spherical surface with radius ${\rm R_{seed}}$. 

Moving forward, we estimated the optical depth $\tau$ of the Comptonising cloud using the relation provided by \cite{zdziarski_96}:
\begin{equation*}
\Gamma= \left[\dfrac{9}{4} + \dfrac{1}{\tau \left(1+ \frac{\tau}{3} \right) \frac{kT_e}{m_ec^2}} \right]^ {1/2} - \dfrac{1}{2},
\end{equation*} 
finding that $\tau = 9.8 \pm 0.2$ and $\tau = 9.6 \pm 0.02$ for NB and FB, respectively. The electron corona is then optically thick and it is likely responsible for the shielding of the emission coming from the innermost region of the system, as hypothesised for other NS-LMXBs \citep[see e.g.][and references therein]{Mazzola_19,iaria_19}.

In both NB and FB spectrum, a broad emission line with energy $E_{\rm line_{\rm K\alpha}} \sim$6.6 keV and width $\sigma_{\rm K\alpha}$ around 0.35 keV was detected; it corresponds to the fluorescence of the \ion{Fe}{xxv} and it is compatible with the results obtained by \cite{dai_07} for NB/FB spectra. We observed also an emission line around 7.8 keV, corresponding to the K$\beta$ transition of the \ion{Fe}{xxv} ion. 
 We infer an equivalent width of $56^{+6}_{-7}$ eV and $52^{+5}_{-6}$ eV for the Fe-K$\alpha$ line in NB and FB, respectively; while the Fe-K$\beta$ emission line shows an equivalent width of $17 \pm 4$ eV in NB and $9^{+4}_{-5}$ eV in FB. These features are compatible with the two \textit{intercombination} lines belonging to the He-like triplets of \ion{Fe}{xxv} \citep[see e.g.][]{Iaria_05}; due to the \textit{NuSTAR} energy resolution, the \textit{forbidden} and \textit{resonance} lines of the triplets are probably blended to the intercombination, which results predominant.
 We estimated the branching ratio $K_{\beta}/K_{\alpha}$ between the intensity of the detected lines as the ratio between the normalisation values of the  corresponding Gaussian components obtained from the best-fit parameters, finding that it is $0.20 \pm 0.04$ and $0.11 \pm 0.04$ in NB and FB, respectively; the uncertainties are calculated at 68\% confidence level. These values are compatible each other within 3$\sigma$ and with the theoretical value of 0.27 (expected for the resonance lines)\footnote{See the NIST Atomic Spectra Database Lines Data, \url{https://www.nist.gov/pml/atomic-spectra-database}}, precisely within 2$\sigma$ for the NB spectrum and within about 3$\sigma$ for the FB spectrum, respectively. The difference between the $K_{\beta}/K_{\alpha}$ values in NB and FB is due to the reduced intensity of the $K_{\beta}$ line in the flaring branch. This one might be related to the different statistics of the data between the two spectra; we infer, indeed, a total number of photons of $\sim 3.6 \times 10^6$ and $\sim 2.51 \times 10^6$ for NB and FB, respectively).
 Unfortunately, no other information about the plasma diagnostic can be obtained due to the limited spectral resolution of the data for the emission lines. Further observations with higher spectral resolution in the Fe-K region of the spectrum are necessary to this purpose.

From model 3B, under the same assumption discussed above, we inferred an inner radius of the accretion disc $R_{\rm disc} = 266^{+104}_{-83}$ km and $107 \pm 21$ km for NB and FB, respectively.
In order to convert this value into the realistic inner disc radius $R_{\rm real}$, we used the relation $R_{\rm {real}} \sim f^2 R_{\rm disc}$, where $f \sim 1.7$ is the colour correction factor and it depends on the mass accretion rate $\dot{M}$ \citep[see][for details]{shimura_95}. We find that $R_{\rm real}= 769^{+301}_{-240}$ km and $310 \pm 61$ km in NB and FB, respectively, which indicates we are not able to observe the innermost region of the disc, and it is compatible with the results obtained from Model 3A. A possible explanation could be provided taking into account the high value of the mass accretion rate $\dot{M}$, that is around $\sim 2.5 \dot{M}_{\rm Edd}$ (for a NS with mass of 1.4 M$_{\odot}$ and radius of 10 km). For such values of $\dot{M}$, the inner regions of the accretion disc are thermally unstable and they result radiation pressure dominated; this inflates the inner disc, generating an optically thick bulge of matter which may act as a shielding corona for the inner soft emission \citep{shakura_76,sincell_97}.

Furthermore, we estimated the optical depth $\tau$ of the Comptonising cloud using the relation described above, finding that $\tau = 7.58 \pm 0.09$ in NB and $\tau = 6.2 \pm 0.3$ in FB. Also in this case, the electron corona is optically thick and the value of $\tau$ is compatible with those obtained by \cite{disalvo_06} in FB and by \cite{dai_07} in NB/FB.
Under this scenario, we estimated also the average electron density $N_e$ of the cloud using the relation $\tau=N_e \sigma_T R$, in which $\tau$ is the optical depth, $\sigma_T$ is the Thomson cross-section and $R=R_{\rm real}-R_{\rm NS}$ is the geometrical dimension of the Comptonising corona, assuming that it covers the innermost region of the system from the NS surface up to the inner radius of the accretion disc. Thus, we obtained $N_e=1.48 \times 10^{17}$ cm$^{-3}$ and $N_e=3.01 \times 10^{17}$ cm$^{-3}$ for $R_{\rm NS}=$10 km, under the assumption that $R_{\rm {real}}$ is the real inner radius of the accretion disc.
Furthermore, we also estimated the radius $R_{\rm seed}$ from which the seed photons are emitted using the relation provided by \cite{zand_99}; we find $R_{\rm seed}=45 \pm 5$ km in NB and $23 \pm 3$ km in FB, suggesting the seed photons were mainly emitted from a region near the NS surface.

Lastly, also in this case we observed two broadened Gaussian lines in the Fe-K region of both NB and FB spectrum around 6.6 keV and 7.8 keV, corresponding to the Fe-K$\alpha$ and Fe-K$\beta$ transition of the \ion{Fe}{xv}, respectively. We infer an equivalent width of $56^{+6}_{-7}$ eV and $52^{+5}_{-6}$ eV for the Fe-K$\alpha$ line in NB and FB, respectively; while the Fe-K$\beta$ emission line shows an equivalent width of $21 \pm 5$ eV in NB and $6^{+5}_{-4}$ eV in FB. The branching ration $K_{\alpha}/K_{\beta}$ is $0.24 \pm 0.04$ in NB and $0.10 \pm 0.05$ in FB, with uncertainties calculated at 68\% c.l., and also in these case these values are compatible with each other within 3$\sigma$ and with the theoretical value within 1$\sigma$ for the NB spectrum and within  3$\sigma$ for the FB spectrum.

As mentioned in the previous section, the reflection component does not represent an improvement for the fit, as well as the smearing component is unnecessary to describe these features, suggesting that reflection occurs at such a distance from the NS surface that the relativistic and Doppler effect are not ruling. 
This hypothesis is endorsed by the larger value inferred from the {\sf diskbb} model for the inner radius of the accretion disc, which would locate the radius at which the reflection originate at more than 250 km; besides, since the source is ultra-bright, it is possible that the inner region of the accretion disc are so ionised for the irradiation due to incident emission from NS/boundary layer to be mainly composed of neutral matter thus preventing the formation of discrete features in the reflection component, as suggested by \cite{homan_18} for the Z-source GX 5-1.
However, it is even possible that the lack of coverage of the data at the lower energy leads to an inefficiency of the self-consistent model into the description of the reflection continuum. Further investigations on a broader energy band are required to shade light about these hypothesis.

In the end, a weak hard tail above 30 keV was detected; it contributes to less than 5\% to the total unabsorbed luminosity in the 0.1-100 keV energy range, both in NB and FB, in disagreement with the contribution of the power-law around 10\%-12\% of the total luminosity observed by \cite{disalvo_06,dai_07,Ding_21} in the 2-200 keV energy range, who found a correlation between the intensity of the power-law and the position of the source in CD. This correlation was demonstrated also by \cite{revnivtsev_14} studying the variation of the amplitude of this component along the Z-track and the relation of the hard tail with the other spectral components. A similar analysis is not possible in our case due to the weakness of the power-law, probably related to the lack of coverage above 60 keV. On the other hand, although the photon index $\Gamma$ results within the range 1.9-3 determined by \cite{dai_07,Ding_21} in NB, it seems to be out of the range 0.2-0.8 identified by the same authors for the FB. However, the uncertainties on $\Gamma$ are large here, resulting actually unconstrained and then hampering the possibility to evaluate the variation of the power-law along the Z-track. The lack of a high-energy cut-off suggests a non-thermal origin for the hard tail \citep[see e.g.][ and reference therein]{disalvo_06}, which could be the result of synchrotron emission from energetic electrons \citep{riegel_70} or a Comptonisation of the soft seed photons on non-thermal electrons \citep{revnivtsev_14} or a non-thermal Comptonisation due to a bulk motion in the nearest of the NS \citep{farinelli_08,Ding_21}, or again as the result of Compton up-scattering of the soft photons in the relativistic radio jets \citep{Reig_16,Reig_15}.
In order to distinguish between these models and find the correlation between the X-ray hard tail and the other spectral components, broad band observations with good statistics along the complete Z-track of the source are needed, especially with simultaneous radio coverage, to eventually correlate it with strong outflows or jets.

\section{Conclusions}
We analysed the 3-60 keV spectrum of Sco X-1 collected by \textit{NUSTAR} in 2014; from the CD of this observation, we separated the spectrum of the normal branch and the flaring branch.

We fitted both the spectra using a model composed of a blackbody component, a thermal Comptonisation and a power-law component, absorbed by the ISM at the lower energies.
Alternatively, a good fit was represented by a model in which the thermal component is provided by a multi-colour blackbody.

In both the cases, two Compton broadened emission lines were detected at 6.6 keV and 7.8 keV corresponding to the the K$\alpha$ and K$\beta$ transition of He-like \ion{Fe}{xxv} ion.
The detection of the Fe k$\alpha$ line is significant in both the CD branches, while the Fe K$\beta$ line seems to be less important in the flaring branch.
A reflection component is not required to model these features.

The presence a power-law with a photon index between 2 and 3 is required for both the models, despite the associated parameters are not well-constrained. The power-law is very weak and contributes to the total flux for $\lesssim$1\% in both the branches.

The two models depict equivalent physical scenarios for the source, with a softening of the spectrum between the normal and the flaring branch. 

\section*{Acknowledgements}
This research has made use of data and/or software provided by the High Energy Astrophysics Science Archive Research Center (HEASARC), which is a service of the Astrophysics Science Division at NASA/GSFC and the High Energy Astrophysics Division of the Smithsonian Astrophysical Observatory.
The authors thank B.W. Grefenstette of NuSTAR SOC for the kind and helpful discussion about NuSTAR data reduction procedures.\\
The authors acknowledge financial contribution from the agreement
ASI-INAF n.2017-14-H.0, from INAF mainstream (PI: T. Belloni; PI: A. De Rosa)
and from the HERMES project financed by the Italian Space
Agency (ASI) Agreement n. 2016/13 U.O.
RI and TDS acknowledge the research
grant iPeska (PI: Andrea Possenti) funded under the INAF
national call Prin-SKA/CTA approved with the Presidential Decree
70/2016. 
S.M. Mazzola extends his utmost gratitude to the editor S. Campana and the anonymous Referee for their useful contribution and their patience during a prolonged refereeing process, due to an unforeseen personal issue of the author herself.

\bibliographystyle{aa}
\bibliography{biblio}
\begin{appendix}
\section{Testing the hard tail significance via Monte-Carlo simulations}
\label{appendix}
We used an alternative method to the F-test probability to evaluate the significance of the presence of the hard tail, employing the Monte-Carlo (MC) technique to simulate a set of 1000 spectra for each \textit{NUSTAR} instrument.

Assuming as null hypothesis that the spectrum does not require a power-law component, we used the corresponding best-fit model (Model 2A) and the {\tt fakeit} task of {\sc XSPEC} for the simulations, obtaining 1000 spectra for both FPMA and FPMB. We used the \textit{NuSTAR} FPMA and FPMB ancillary response file (ARF), response matrix (RMF) and background spectrum for each instrument. We simulated each spectrum for the same amount of exposure time both for the source and background.
Each couple of simulated FPMA+FPMB spectra was fitted using both the best-fit model 3A and 2A (including and not-including the power-law component, respectively). For each fit, we evaluated the discrepancy $\Delta \chi^2$ between the two obtained $\chi^2$. Each value of the obtained  $\Delta \chi^2$ was compared with the same value of $\Delta \chi^2$ obtained from the real data.

We evaluated the distribution of the $\Delta \chi^2$ reporting the number of times in which the simulated value of $\Delta \chi^2$ is larger than that obtained with the real data (i.e. good trials). In the end, the probability of chance improvement has been evaluated by considering the good trials value divided by the total number of simulations.

We followed the described approach for both the branches.
For the NB, we obtained a $\Delta \chi^2$ larger than the real one 0/1000 times, then the probability of chance improvement is $<0.1\%$.  While for the FB, the obtained $\Delta \chi^2$ results larger than the real one 9/1000 times, obtaining a probability of 0.9\%. In both the cases, the simulations confirm what predicted by the F-test.

We applied the same procedure to the B-models, simulating the spectra with Model 2B and fitting them with Model 2B and 3B. We obtained a probability of chance improvement $<0.1\%$ in NB and  of 0.7\% in FB, confirming the F-test prediction also in these cases.

By relying on the MC tests, we can conclude that the power-law component is required in both the branches with the confidence level reported in the text.
\end{appendix}
\end{document}